\documentclass[12pt,manuscript]{aastex}
 \usepackage{txfonts}
 \usepackage{graphicx}
 \usepackage{epstopdf}
 \usepackage{natbib}
\slugcomment{For submission to the Astrophysical Journal Letters}
\shorttitle{Sunspot and Starspot Lifetimes}
 \shortauthors{Bradshaw and Hartigan}
 
\begin{document}
\tighten
 \title{On Sunspot and Starspot Lifetimes}
 
 \author{S. J. Bradshaw}
 \affil{Department of Physics and Astronomy, Rice University, Houston, TX 77005, USA}
\email{stephen.bradshaw@rice.edu}\and

 \author{P. Hartigan}
\affil{Department of Physics and Astronomy, Rice University, Houston, TX 77005, USA}
 \email{hartigan@sparky.rice.edu}
 

\begin{abstract}
We consider the lifetimes of spots on the Sun and other stars from the standpoint
of magnetic diffusion.  While normal magnetic diffusivity predicts lifetimes of
sunspots that are too large by at least two orders of magnitude, turbulent magnetic
diffusivity accounts for both the functional form of the solar empirical
spot-lifetime relation and for the observed sunspot lifetimes, provided that the
relevant diffusion length is the supergranule size. Applying this relation to
other stars, the value of turbulent diffusivity depends almost entirely on
supergranule size, with very weak dependence on other variables such as magnetic
field strength and density. Overall, the best observational data for other stars
is consistent with the extension of the solar relation provided that stellar
supergranule sizes for some stars are significantly larger than they are on the Sun. 
\end{abstract}

 

 \section{Introduction}
 \label{introduction}

Starspots are ubiquitous features of stellar surfaces and markers of stellar
activity, yet we know relatively little about the physical processes that govern
their lifetimes. Spot lifetimes are important to quantify because (i) if lifetimes
are long enough, stellar lightcurves from different epochs may be phased over
multiyear timescales, which facilitates rotation and angular momentum studies of
young stars; (ii) measurements of spot lifetimes allow stellar activity cycles
to be tracked and studied; and (iii) lifetime measurements yield insights into
the physics of stellar convection zones. Sunspot lifetimes may also provide clues
to the time dependence of the heating mechanism in solar active regions.
For example, \citet{Ugarte-Urra2012} found
that young active regions are consistent with more dynamic, burst-like low-frequency
heating, whereas the energy deposition in older active regions may have evolved
to a relatively steady-state configuration corresponding to high frequency heating.
 
An obvious question to consider is whether commonalities exist between the natures
of sunspots and starspots, and there is some evidence that they do. We possess a
certain amount of phenomenological information based on long-term photometric
observations of stellar brightness variations; for example, the lifetimes of
smaller spots are proportional to their sizes \citep{Burdyugina2005}, which is
also true of sunspots \citep{Gnevyshev1938,Waldmeier1955,Petrovay1997}. In
addition, long-term monitoring of stellar lightcurves has revealed what appear
to be the analog of solar cycles that operate on timescales of decades \citep{olah09}.
Like the Sun, spotted stars are also well-known to have enhanced X-ray activity
associated with magnetic coronal heating \citep[e.g.][]{feigelson05}.

Significant differences between sunspots and starspots also exist. For example,
sunspots typically last for days to a month, and cover at most a few percent of
the Sun's surface area, while spots on young, pre-main-sequence, solar mass
(T-Tauri) stars may persist for several years and cover up to 20\% of the surface
\citep{Hall1994,Strassmeier1999b}. There is a greater
brightness contrast between spot regions and the surrounding photosphere on hotter
stars, which suggests that the temperature difference between them increases with
stellar temperature.  The temperature difference is about 2000~K in stars of
spectral type G0 and only 200~K in stars of type M4. This property persists
between active dwarfs and giant stars, where cooler dwarfs tend to have stronger
magnetic fields covering larger areas \citep{Burdyugina2005}. Another difference
between the solar and stellar cases is that some stars exhibit long-lived polar
spots \citep{Strassmeier1999b}. 

One stellar property that might act to limit spot sizes and hence their lifetimes
is differential rotation, which causes shearing that breaks up large spots that
cover a broad latitudinal range into fragments that subsequently form groups of
smaller spots with correspondingly shorter lifetimes. \cite{Hall1990} studied the
lifetimes of a sample of 40 spots on 17 different stars and found that small spots
($r_s < 20^\circ$) disappear before they are subjected to significant shearing,
whereas larger spots appear to be more disrupted by differential rotation. The
rotational shear is defined as the difference in the rate of rotation between the
equator and the pole, given by $\Delta \Omega$. In solar-type differential
rotation the equatorial region rotates faster than the poles with
$\Delta \Omega = 0.055$~rad~day$^{-1}$. The strength of the differential rotation
is quantified by $\tau_L = 2 \pi / \Delta \Omega$~days, where $\tau_L$
($\sim$ 115 days for the Sun) is the time taken for the equator to lap the poles.
Clearly, if $\tau_L < \tau_S$, where $\tau_S$ is the spot lifetime, then
differential rotation is likely to be important in determining the longevity of
the spot. In the case of the Sun $\tau_L$ is significantly more than the longest
observed spot lifetime of approximately 1~month ($\approx 30$~days), so the
differential rotation is probably not a factor.

Differential rotation can be identified by long-term photometric observations of
the stellar brightness modulation, where changes in the seasonal rotation period
are indicative of differential rotation and the gradual migration of spot latitudes
\citep{Henry1995}. A comparison between the range of seasonal rotation rates and
the mean rotation rate of the star suggests that slower rotators exhibit stronger
differential rotation. Significantly, a majority of stars are found to have weaker
rotational shear than is observed on the Sun. This was confirmed by
\cite{Reiners2003a,Reiners2003b} who used a Fourier transform method to study the
differential rotation of a sample of rapidly rotating F0 - G0 dwarf stars and
found that it is strongest in slower rotators. Moreover, the differential rotation
of the more rapidly-rotating stars in their sample fell below their method's
threshold of sensitivity, rendering it undetectable. The differential rotation
of a small sample of active G2 - M2 dwarfs was analyzed by cross-correlating
successive Doppler images \citep{Petit2004,Barnes2005}, a more sensitive
technique than the previous Fourier transform method, which revealed that
$\Delta \Omega$ is effectively negligible in M dwarfs and strong in G dwarfs.
A particularly large brightness variation (0.65 mag) was observed for V410 Tau,
where the lifetime of the associated spot has been estimated as at least 20 years
\citep{Strassmeier1997,Hatzes1995} indicating that differential rotation may not
play a significant role, especially if, like V410 Tau, the star is a relatively
fast rotator.

In light of the possibility that the underlying nature of starspots and the
physical processes determining their lifetimes may be similar from star to star,
in this paper we investigate whether the proportional relationship between the
lifetime of a spot and its size is consistent with anomalous magnetic diffusivity
operating at the supergranule scale. In Section~\ref{TheGWRule} we discuss the
encapsulation of this relationship by the Gnevyshev-Waldmeier Rule and how it may
be applied to the Sun and T-Tauri stars. In Section~\ref{MD} we discuss the
reasons why classical magnetic diffusivity cannot account for the observed spot
lifetimes, and in Section~\ref{AMD} we introduce the need for anomalous magnetic
diffusivity to explain the relative lifetimes of spots on the Sun and other stars.
We consider what is known about spot lifetimes for the best-observed cases in
Section~\ref{obs} , and summarize the paper in Section~\ref{SandC}.
 
\section{The Gnevyshev-Waldmeier Rule}
\label{TheGWRule}

The Gnevyshev-Waldmeier rule \citep[][hereafter G-W]{Gnevyshev1938,Waldmeier1955}
is a simple linear empirical relationship between the area of a sunspot and its lifetime,
 
\begin{equation}
\label{eqn1}
A = WT
\end{equation}
 
\noindent
where $A$ is the maximum size of the spot, $T$ is the spot lifetime and
$W=10$~MSH~day$^{-1}$ for the Sun. The unit of area {\it Micro Solar Hemisphere}~(MSH)
is $\left(5.8^{\arcsec}\right)^2$, where $1^{\arcsec} \approx 725$~km on the Sun.
Hence, the rate of area shrinkage of the spot, $W=3.04 \times 10^{13}$~m$^2$~day$^{-1}$
is the quantity encapsulating the physics that determines the lifetime of the spot
and is of direct interest to us in the present work. Many sunspots persist for
less than one day, and lifetimes of more than a week or two are relatively
uncertain since the spots disappear over the limb and observations are interrupted.
The data used to calculate $W$ are probably biased toward smaller values of $T$
than are obtained for the longest-lived sunspots.
 
Applying the G-W rule to a spot which might be typical of a T-Tauri star can yield
an estimate of the spot's longevity, which we expect to be significantly extended in
comparison with the average lifetime of a sunspot. A typical T-Tauri star has a
radius of $3R_\odot$ and lightcurve amplitudes that range from a few percent to
ten percent. The total lightcurve amplitude will depend on the relative amounts
of starspots on opposite sides of the star, but for the purpose of calculation
if we take an individual spot to cause a light variation of 0.01 mag and assume
the spot to be completely dark, then the spot area is $1.4 \times 10^{17}$~m$^2$,
which corresponds to a lifetime of $\approx 12$~years using Equation~\ref{eqn1}.
These lifetime estimates are roughly in line with what we expect based on
observational studies of T-Tauri stars; that is, the relationship between the
sizes and lifetimes of spots provides values that fall within the right orders of magnitude.
We will now consider the physics that underlies the value of $W$, and assess if
there are theoretical reasons why
the solar relation should hold for active stars in general.
 
\section{Magnetic Diffusivity}
\label{MD}

Our aim is to identify the physical nature of the sunspot diffusion so that we
can know how to scale it and apply it to other stars.
Spots on active stars are regions of relatively strong magnetic field, so we begin
our investigation by determining whether their observed lifetimes are consistent
with classical magnetic diffusivity. The rate of change of field strength in the
diffusive limit is given by the induction equation,
 
\begin{equation}
\label{eqn2}
\frac{\partial B}{\partial t} = \eta \nabla^2 B,
\end{equation}
 
\noindent
where $\eta$ is the magnetic diffusivity. The G-W rule implies a single
value for the magnetic diffusivity $-$ this would be the actual diffusivity that
operates in the Sun and does not depend on the spot size. To order of magnitude,
the lifetime of a spot due solely to the rate of diffusion of its magnetic field is given by
 
\begin{equation}
\label{eqn3}
\tau = \frac{{R^2_{spot}}}{\eta},
\end{equation}

\noindent
where $R_{spot}$ is the radius of the spot.
The classical magnetic diffusivity is the inverse of the product of the magnetic
permeability $\mu_0$ and the electrical conductivity $\sigma$ due only to Coulomb
collisions.
 
\begin{equation}
\label{eqn4}
\eta = \frac{1}{\mu_0 \sigma},
\end{equation}
 
\noindent
where $\mu_0 = 4\pi \times 10^7$~N~A$^2$. It is straightforward to
calculate the electrical conductivity in the case of a fully ionized plasma, but
estimating $\sigma$ in the partially ionized photosphere is more difficult and
its value depends to some extent on the model
atmosphere on which the calculation is based. \cite{Kovitya1983} calculated the
conductivity using a second-order Chapman-Enskog method \citep[which differs from
the fourth order approximation by less than 0.2\%; ][]{Devoto1967} and the sunspot
model of \cite{Avrett1982}. The ionization-equilibrium calculation used by
\cite{Kovitya1983} predicted a somewhat weaker charge state (lower electron
density) than in the original sunspot model, and a slightly lower value for
$\sigma$. To be conservative, we choose the smallest value of $\sigma$ in the
coolest part of the sunspot, which in Table~II of \cite{Kovitya1983}
is $\sigma=1.3$~S~m$^{-1}$. Inserting this quantity into Equation~\ref{eqn4} we
find that $\eta = 5.29 \times 10^{10}$~m$^2$~day$^{-1}$, much smaller than
the G-W value ($\eta_{GW} \sim W/\pi$) of $\eta_{GW}=9.68 \times 10^{12}$~m$^2$~day$^{-1}$
($\eta_{GW} / \eta \approx 180$). Larger values of $\sigma$ make the comparison even worse.

Table~1 shows the sunspot and starspot lifetimes expected for a given spatial
scale based on the classical magnetic diffusivity and the electrical conductivity
in the photosphere. It is clear that the lifetimes are gross overestimates
(by $\gtrsim$ 2 orders of magnitude).
The electrical conductivity depends on the composition of
the photosphere and the dominant term in calculating its value is the elastic
scattering of electrons from neutral hydrogen \citep{Kovitya1983}. In consequence,
provided that the photosphere is composed primarily of hydrogen and spot interior
temperatures are not hugely variable then there should not be large
variations in $\sigma$ from star to star.
 
\section{Anomalous Magnetic Diffusivity}
\label{AMD}

Because classical magnetic diffusivity based on Coulomb collisions is not of
sufficient magnitude to explain the lifetimes of observed spots, we must appeal
to a form of anomalous diffusivity if magnetic diffusion governs spot lifetimes.
\cite{Chae2008} have suggested that the magnetic diffusivity is of turbulent
origin and this should be accounted for in the rate of flux cancellation at the
photosphere. They measured values of the diffusivity by applying the induction
equation to pairs of magnetograms taken at different times by {\it Hinode}~SOT
and {\it SOHO}~MDI at pixel sizes of 116~km (SOT), 440~km (high-resolution MDI
magnetograms) and 1400~km (full disk MDI magnetograms). By solving the induction
equation for $\eta$ at these different length scales, they were able to estimate
values for the magnetic diffusivity and found them to be consistent with a
turbulent cascade from supergranular and granular scales, ending at a resistive
dissipation scale of about 30~km (the scale at which the magnetic diffusivity
equals the classical collisional value).

Table~2 provides some estimates of sunspot and starspot lifetimes based on the
\cite{Chae2008} values of the anomalous magnetic diffusivity.  The lifetimes
predicted for sunspots and starspots in Table~2 are still too long at the scales
listed in the first column. We may then ask: what is the length scale of
$\eta$ consistent with the G-W value? \citet{Chae2008} show that the observed
cascade follows the Iroshnikov-Kraichnan model for magnetohydrodynamic turbulence,

\begin{equation}
\label{eqn11}
\eta = \left( \frac{l}{260~\mbox{km}} \right)^{5/4}~~~~\mbox{km$^2$~s$^{-1}$}.
\end{equation}
 
\noindent
where $l$ is the turbulent scale length. We match the G-W value
($\eta_{GW} = 112$~km$^2$~s$^{-1}$) for $l$ = 11300~km, which is the size
of a typical supergranule \citep{Leighton1962,Chae2008}.

Thus, turbulent magnetic diffusivity at supergranule size scales that cascades
down to a scale at which collisional diffusivity operates and energy is dissipated
appears to provide a viable explanation for the lifetimes of sunspots and T-Tauri
starspots. Other authors \citep[e.g. ][]{Petrovay1997b} have
found that magnetic diffusion operating on the granular scale ($l \approx 1500$~km)
is adequate to explain observed sunspot lifetimes. However, this conclusion
results from using a much larger value for $\eta$. \cite{Petrovay1997b} quote a granular
value of $\eta = 1000$~km$^2$~s$^{-1}$ (which is a factor of $\sim$ 9 larger than
$\eta_{GW}$) in comparison to the later measurements of \cite{Chae2008}, who found
that $\eta=18$~km$^2$~s$^{-1}$ at $l=1400$~km (see also Table~II).

In the Iroshnikov-Kraichnan model, the turbulent magnetic diffusion is given by
 
\begin{equation}
\label{eqn12}
\eta = \left( \epsilon V_A \right)^{1/4} l^{5/4},
\end{equation}
 
\noindent
so that the scaling of the diffusion coefficient only depends weakly on the rate
of energy injection $\epsilon$ and the Alfv\'{e}n speed $V_A$. The weak scaling
with the Alfv\'{e}n speed implies a weak scaling with the magnetic field strength
$B$ and the particle number density $n$ ($V_A \propto B / \sqrt{n}$), quantities
that may vary significantly from star to star. This result suggests that as long as the size of supergranules remains the same,
the magnitude of the turbulent diffusivity in other stellar photospheres
could be similar to that of the Sun. Otherwise, one should be able to use the solar G-W law suitably scaled
to the supergranule size, for other stars.

\cite{Leighton1964} was the first to associate the dispersal and migration
of photospheric magnetic regions with a random-walk, diffusive process driven by
supergranulation-scale convection currents. He estimated very similar dispersal
rates due to granular and supergranular convection, but argued that only the more
deeply rooted supergranular motions are coupled strongly enough with the magnetic
field lines to move them about. Evidence for this is the chromospheric network,
which is described by the distribution of magnetic flux that corresponds to the
boundaries of supergranule cells. \cite{Simon1964} suggested that one possible
mechanism for the breakup of sunspots at this scale is the progressive fragmentation
of the magnetic field by the action of the supergranulation convection currents.

Returning again to a classical formalism,
the magnetic diffusivity depends on the length scale for dissipation
$l$ and the gas velocity $u'$ by
\citep[e.g. ][]{Leighton1964}

\begin{equation}
\label{eqn13}
\eta = \frac{1}{3} u' l.
\end{equation}

\noindent
If turbulent flow dominates, then $u'$ should be the magnitude of velocity fluctuations.
The first three lines of Table~II show the observed diffusivities for the size scales
reported by \cite{Chae2008}, and the corresponding velocities inferred from equation~\ref{eqn13}.
These velocities compare well with observations of actual velocity fluctuations in the solar photosphere.
\cite{Lites1998} measured oscillations of Doppler velocity at photospheric heights
in a sunspot based on observational data taken by the High Altitude Observatory
(HAO)~/~National Solar Observatory (NSO) Advanced Stokes Polarimeter (ASP).
This instrument provides the complete Stokes polarization vector profiles of
several spectral lines (e.g. Fe~I 630.15~nm, 630.25 nm) and allows vector magnetic
field and Doppler velocities in sunspots and active regions to be measured to a
high degree of accuracy. Table~I of \cite{Lites1998} reports velocity
fluctuations of $\sim 40-50$~m~s$^{-1}$ in the umbral and penumbral regions in
the 5 minute band (the 3 minute band is chromospheric~/~transition region).
Lifetimes of spots for the scale sizes in the first three lines of Table~II are still much longer
than the observed ones, so for turbulent diffusion to explain the observations
the length scales must be larger than those observed by \cite{Chae2008}.
Using the value of $\eta$ required by the GW law, and taking $l$ to be a
supergranular scale of $l=11300$~km, the velocity given by Equation~\ref{eqn13} is
$u' = 30$~m~s$^{-1}$, reasonably consistent with the observed fluctuations.

However, once the dissipation scale size exceeds the size of the spot, 
diffusion becomes a non-local process and equation~\ref{eqn13} may not apply.
The transport of magnetic flux in a weakly electrically
conducting plasma could be analogous to the transport of
heat in a plasma where the mean-free-path approaches the physical scale of the
system. The temperature and ionization several thousand kilometers below the
photosphere indicate a high electrical conductivity and a slow rate of diffusion,
which may serve to limit the displacement of `free-streaming' field lines at the
surface, causing a localized build-up of flux and then strong dissipation due to
the enhanced field gradient. Alternatively, the scales relevant to determining
the physics that sets the lifetime of the sunspot may not be visible at the solar
surface; the photosphere is only the layer at which the optical depth of light at
$\sim$ 550~nm approaches 2/3, but doesn't have to be the layer at which key physical
scales relevant to magnetic activity are revealed. Finally, it may simply be that
the velocity fluctuations discussed above (usually averaged out in the data),
that are essentially scale-independent up to at least a supergranular scale, lead
to an effective diffusion of magnetic flux concentrations \citep{Litvinenko2011}.

\section{Observational Data and Discussion}
\label{obs}

While a large amount of evidence for spots on other stars exists, usually in the
form of small-amplitude periodic variations in the stellar lightcurves, it can be
very difficult to infer the sizes and lifetimes of spots accurately from photometry
alone owing to degeneracies in reconstructed maps of the stellar surfaces
\citep{walk13,lanza13}. However, when combined with the techniques of Doppler
imaging and eclipse mapping by transiting exoplanets, photometric data yields a
much more robust assessment of the size and longevity of starspots, albeit only
for a limited selection of objects.  To augment the solar observations, in Table~3
we compile starspot data from recent papers of three active main-sequence stars
that the Kepler and CoRoT satellites have found to have transiting planets.
These values derive from maps of the starspot distributions that typically
show the spots last for a month or two and cover something like 20 degrees in 
longitude.  In order to sample the largest range of parameters possible, we also include two
other well-known stars that have been reported to host large spots.
This sample is not intended to be complete, but rather to provide an indication
as to how the parameters vary among different stars.

The lifetimes and sizes of starspots on the main-sequence transiting planet objects
Kepler~17, CoRoT~6 and CoRoT~2 \citep[][respectively]{kepler17,corot6,corot2a,corot2b}
in Table~3 resemble scaled-up versions of the Sun.  For the two stars with larger
radii, Doppler imaging of the rapidly-rotating young star V410~Tau  shows a large
spot near the pole that has persisted for at least 20 years \citep{Hatzes1995,v410taub},
and the FK Comae giant HD~119178 has an even larger polar spot that has lasted for
more than 12 years \citep{Strassmeier1999b}. While the origin of these polar spots
may differ from those at lower latitudes, it is clear that, like sunspots,
starspots live longer when they are larger, as expected if a diffusion process
governs their longevity.

Plotting the lifetime of the starspots against their size, we obtain Fig.~1.
The extension of the solar G-W law using a constant supergranule size lies above
the observed points for the solar analogs with transiting planets, but is consistent
with the lower limits of the ages for the large polar spots on the subgiant and
giant stars. Taken at face value, matching the lifetimes of the solar analog stars
requires a supergranule diffusion length much larger than the Sun, covering
$\sim$ 23$^\circ$ in longitude.
However, one should keep in mind that estimates of spot sizes on these objects
are limited by spatial resolution, so the spot sizes could be considerably smaller
than estimated in Table~3 if they collect loosely into larger active areas, and this behavior
would reduce the required supergranule size.  The systematic errorbar of a factor
of two in the figure represents the approximate uncertainty in assigning a
lifetime and a spot size from the reconstructed images of the stellar surfaces.  

Explaining the origin of supergranulation remains one of the outstanding challenges
of modern solar physics \citep{Rieutord2010}. It is not clear what sets the
lifetime and, most importantly for the work presented here, the spatial dimension
of the supergranule cells. We therefore do not have a good idea, or even much
intuition, of how this quantity might vary from star to star. The typical size
scale for supergranulation on the Sun is 10,000 to 30,000~km \citep{Hirzberger2008},
with a depth of not more than 5,000~km \citep{Sekii2007} as determined by local
helioseismology, which suggests that supergranules are in fact rather shallow
structures. Their depth may be linked to the ionization depth of He or
He$^+$ \citep[$2000 - 7000$~km; ][]{Simon1964} in a similar manner to the depth
of granulation cells being connected with the location of hydrogen ionization
below the photosphere. The average size of supergranules is sensitive to the
method used to define them (e.g. as coherent, diverging flow cells at the solar
surface) and their vertical profile is at present unconstrained. One potential
clue to the mechanism that sets the size scale of supergranulation may lie in
observations of its dynamical interaction with magnetic field in the quiet Sun.
Supergranules are strongly associated with the magnetic network and correlations
between their sizes and the strength of network and internetwork fields has been
observed \citep{Meunier2007}.

Results from numerical simulations have been
similarly inconclusive. Global simulations have yielded a supergranulation-like
pattern, but the scale of the pattern was extremely close to the chosen grid
scale \citep{DeRosa2002}. Local simulations of hydrodynamic convection do not
exhibit excess power at supergranulation scales \citep{Stein2009}, but local
simulations of magnetohydrodynamic convection lead to the formation of a magnetic
network at supergranulation scales in which magnetic flux concentrations may play
an important role in the scale selection process \citep{Ustyugov2009} as suggested
by observations of the quiet Sun. 

Lacking a robust physical understanding of what sets the spatial scale of
the supergranulation makes estimating this scale on other stars extremely difficult.
We can calculate the value of $\eta$ required to explain the lifetime of the
observed spots, but without an estimate of the supergranulation scale or the
velocity fluctuations we face a problem of degeneracy in understanding exactly
what determines $\eta$. One approach to estimating $l$ might be to calculate the
interior temperature structure of the star based on observations of its surface
properties and the standard stellar structure equations, in order to find the
depth of He and He$^+$ ionization as a proxy for the depth of a supergranule.
If $R=D/H$, where $D$ is the cell depth and $H$ its width, \cite{Simon1964}
suggested that the depth of a solar supergranule could be estimated from its
horizontal scale through comparisons with other examples of non-stationary
convection found in nature, where $R \sim 5-10$ over a very large range of
densities and sizes. Conversely, one may estimate the horizontal scale of a
stellar supergranule given its depth.

\section{Conclusions}
\label{SandC}

Starspot lifetimes are of great interest to the astronomical community both from
the standpoint of providing a means to measure rotation periods and activity
cycles, and from a more fundamental plasma physics point of view. Motivated by
the recent improvements in stellar tomography made possible by Doppler imaging
and exoplanet transits, we have attempted to explain the emerging correlation of
stellar spot ages and sizes by considering well-known laws that govern the
phenomenon in the Sun.

The solar case provides insight into the physics of spot lifetimes in several ways.
First, it is clear that classical diffusion processes lead to spot lifetimes that
are too large by at least two orders of magnitude. Recent observations of anomalous, turbulent-driven
magnetic diffusion however, are far more promising, and can account for the solar
data if supergranules represent the maximum scale size for diffusion.
Theoretically, the magnetic diffusion law should be relatively insensitive to
changes in photospheric properties such as the density and magnetic field strength,
and should be governed primarily by the supergranule scale length. However,
while extrapolation of the solar law is consistent with the longevity of large
polar spots on giant and subgiant stars, applying the solar law to active
main-sequence stars overestimates the ages of these spots. Increasing the
supergranule~/~diffusion scales on these stars to cover $\sim$ 20 degrees in longitude
brings the observations in accord with theory. \cite{Hagenaar1999}
hypothesized the presence of a much larger diffusive scale associated with
giant cells of ten times the supergranular length scale. It is also possible,
however, that the spot sizes are overestimated on these stars, and the active
regions consist of many smaller spots grouped together, which would then reduce
the supergranule size needed to explain the observations. This would be
the case if, for example, the spots were fragmented due to buffeting by
supergranule-scale convection currents.

A much larger sample of stars will be needed to further work along these lines,
a difficult task because of inherent ambiguities associated with inverting
lightcurves to generate stellar surface maps. Exoplanet transit mapping combined
with Doppler imaging seems to produce the most robust maps, and with the large
efforts underway to study such systems, one of the byproducts of these efforts
should be a greater understanding of the physical processes at work as starspots
grow and dissipate.

\acknowledgements
 

\begin{figure}
\includegraphics[width=\textwidth]{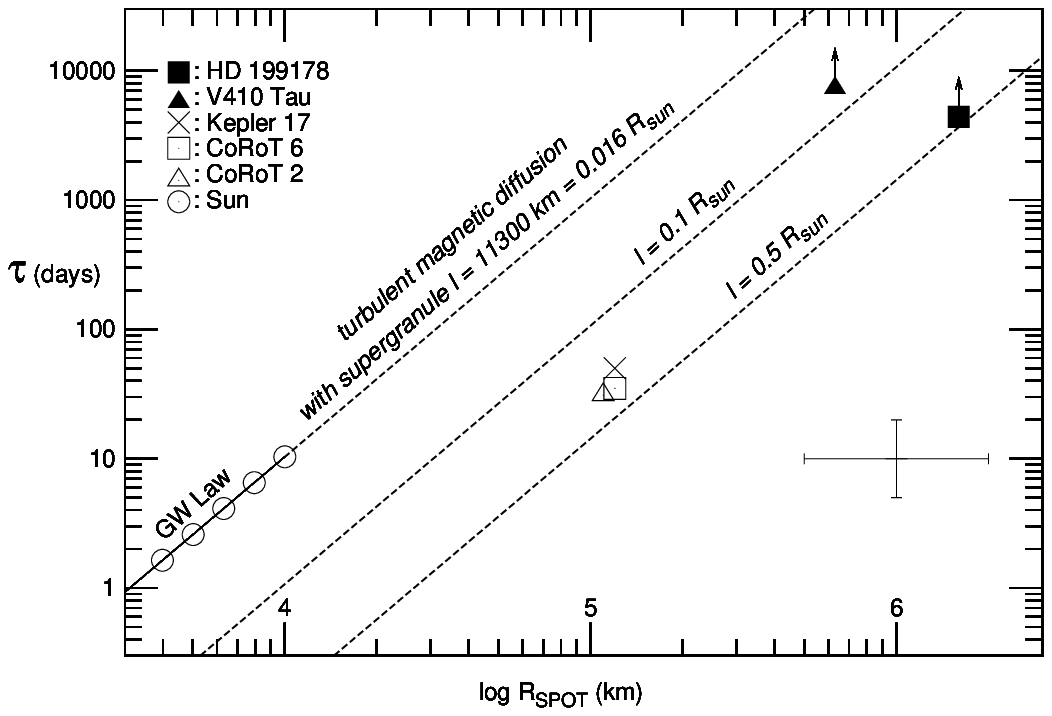}
\caption{Spot lifetime plotted against size. The GW law (solid line) and its extension (dashed line)
are shown for the Sun, and for two cases with larger supergranule sizes $l$. The errorbar represents a
systematic error of a factor of two in both axes.}
\label{fig1}
\end{figure}

\null
\begin{center}
\begin{deluxetable}{ccccc}
\singlespace
\tablenum{1}
\tablewidth{0pt}
\tablecolumns{5}
\tabcolsep = 0.16in
\parindent=0em
\tablecaption{Lifetimes of sunspots and starspots assuming classical magnetic diffusivity}
\startdata
\noalign{\medskip}
\noalign{\medskip}
\noalign{\hrule}
\noalign{\medskip}
Type & \multispan2{\hfil Dimension\hfil } & \multispan2{\hfil Lifetime\hfil } \\
& $\left(R_\odot\right)$ & (km) & (days) & (years) \\
\noalign{\smallskip}
\noalign{\hrule}
\noalign{\smallskip}
Sunspot & 0.04 & $2.8 \times 10^4$ & $1.5 \times 10^4$ & 41 \\
Starspot & 0.33 & $2.3 \times 10^5$ & $1.0 \times 10^6$ & 2700 \\
\enddata
\end{deluxetable}
\label{table1}
\end{center}
 
\begin{center}
\begin{deluxetable}{cccccc}
\singlespace
\tablenum{2}
\tablewidth{0pt}
\tablecolumns{5}
\tabcolsep = 0.04in
\parindent=0em
\tablecaption{Spot lifetimes for the Sun and a T Tauri star assuming turbulent magnetic diffusion }
\startdata
\noalign{\medskip}
\noalign{\medskip}
\noalign{\hrule}
\noalign{\medskip}
Size Scales$^a$ (km) & $\eta$~(km$^2$~s$^{-1}$) & $\eta$~(m$^2$~day$^{-1}$) & $u'$ (m~s$^{-1}$) & $\tau_\odot^b$~(years) & $\tau^c_{\mbox{star}}$~(years) \\
\noalign{\smallskip}
\noalign{\hrule}
\noalign{\smallskip}
116 & 0.87 & $7.5 \times 10^{10}$ & 23 & 28   & 1900 \\
440   & 4.40 & $3.8 \times 10^{11}$ & 30 & 5.5 & 380 \\
1400 & 18.0 & $1.6 \times 10^{12}$ & 39 & 1.4  & 93 \\
11300 & 112 & $9.68 \times 10^{12}$ & 30 & 0.22 & 15 \\
\enddata
\tablenotetext{a:}{Maximum turbulent length scale}
\tablenotetext{b:}{For a sunspot with R$_{\mbox{spot}}$=0.04 R$_\odot$ and the turbulent scale size in column 1,
$\tau$ = R$_{\mbox{spot}}^2$ / $\eta$.}
\tablenotetext{c:}{For a starspot with R$_{\mbox{spot}}$=0.33 R$_\odot$ and the turbulent scale size in column 1}
\end{deluxetable}
\label{table2}
\end{center}
 
\null
\begin{center}
\begin{deluxetable}{lccccccc}
\singlespace
\tablenum{3}
\tablewidth{0pt}
\tablecolumns{4}
\tabcolsep = 0.04in
\parindent=0em
\tablecaption{Observed Starspot Sizes and Lifetimes}
\startdata
\noalign{\medskip}
\noalign{\medskip}
\noalign{\hrule}
\noalign{\medskip}
Star&Spectral Type & P$_{\mbox{rot}}$& R$_{\mbox{star}}$& Spot Radius& Lifetime&References\\
& & (days)& (R$_\odot$)& ($10^5$km)& (days)&\\
\noalign{\smallskip}
\noalign{\hrule}
\noalign{\smallskip}
Kepler 17& G2V  &11.89& 1.05& 1.2 & 50 &a \\ 
CoRoT 6  & F9V  &8.88&  1.03& 1.2 & 35 &b \\
CoRoT 2  & G7V  &4.54&  0.90& 1.1 & 31 &c \\
V410 Tau & K4IV &1.87&  3.46& 6.3 &$>$7300 &d \\ 
HD 199178& G5III-IV&3.32& 6.5& 16 &$>$4400 &e \\ 
Sun      & G2V  &25.4&  1.00& 0.06 & 3.7 & \\
\enddata
\tablenotetext{a:}{ \citet{kepler17}}
\tablenotetext{b:}{ \citet{corot6}}
\tablenotetext{c:}{ \citet{corot2a,corot2b}}
\tablenotetext{d:}{ \citet{Hatzes1995,v410taub}}
\tablenotetext{e:}{ \citet{Strassmeier1999b}}
\end{deluxetable}
\label{table3}
\end{center}

\end{document}